\documentclass[aip,apl,reprint,twocolumn,graphicx]{revtex4-1}
\usepackage{graphicx}
\usepackage{amsmath, amsthm, amssymb}
\usepackage[usenames,dvipsnames]{xcolor}

\def\benum#1\eenum{\begin{enumerate}#1\end{enumerate}}
\def\be#1\ee{\begin{equation}#1\end{equation}}
\def\bml#1\eml{\begin{multline}#1\end{multline}}
\def\ba#1\ea{\begin{align}#1\end{align}}
\def\bas#1\eas{\begin{align*}#1\end{align*}}
\def\eq#1{(\ref{eq:#1})}

\def\bit{\begin{itemize}}
\def\eit{\end{itemize}}

\def\bg#1\eg{\begin{gather}#1\end{gather}}

\def\bm#1\em{\begin{pmatrix}#1\end{pmatrix}}
\def\bg#1\eg{\begin{gather}#1\end{gather}}
\def\t{\text}

\begin{document}

\title{Characterization of low-temperature microwave loss of thin aluminum oxide formed by plasma oxidation}%

\author{Chunqing Deng}
\email{cdeng@uwaterloo.ca}
\author{M. Otto}
\author{A. Lupascu}
\email{alupascu@uwaterloo.ca}
\affiliation{Institute for Quantum Computing, Department of Physics and Astronomy, and Waterloo Institute for Nanotechnology, University of Waterloo, Waterloo, Ontario N2L 3G1, Canada}

\date{\today}

\begin{abstract}
We report on the characterization of microwave loss of thin aluminum oxide films at low temperatures using superconducting lumped resonators. The oxide films are fabricated using plasma oxidation of aluminum and have a thickness of 5~nm. We measure the dielectric loss versus microwave power for resonators with frequencies in the GHz range at temperatures from 54 to 303~mK. The power and temperature dependence of the loss is consistent with the tunneling two-level system theory. These results are relevant to understanding decoherence in superconducting quantum devices. The obtained oxide films are thin and robust, making them suitable for capacitors in compact microwave resonators.

\end{abstract}

\maketitle

Energy loss at microwave frequencies is one of the key aspects of the physics of amorphous dielectrics, a central topic in condensed matter research~\cite{phillips2011amorphous}. In recent years, renewed interest in this topic emerged in the field of superconducting quantum devices\cite{Devoret2013}. These devices unavoidably contain amorphous dielectrics, which are a significant source of decoherence~\cite{Martinis2005, Gao2008, Barends2008, Macha2010, Paik2010}. Reduction of decoherence due to amorphous dielectrics can be achieved by their partial elimination, using suitably modified microwave circuit designs~\cite{Megrant2012,Geerlings2012,Cicak2010,Weber2011}. However, these designs come at the cost of larger space and more difficult fabrication processes. Since the partial elimination of amorphous dielectrics has drawbacks, it is very relevant to investigate the physics of dielectric loss with the aim of reducing the intrinsic loss in these materials.

In this letter, we present results on the fabrication and detailed characterization of microwave loss in thin films of amorphous aluminum oxide. This material is highly relevant in superconducting devices, since it forms the tunnel barrier in Josephson junctions. The films, obtained by plasma oxidation of a deposited aluminum layer, have a thickness of 5~nm. We fabricate overlap capacitors formed by the oxidized aluminum layer and a second aluminum layer. Microwave loss is measured based on lumped resonators which contain these overlap capacitors. Our characterization method is complementary to work on two-level spectroscopy using different types of superconducting qubits~\cite{Martinis2005, Kim2008, Lupaifmmodemboxcselsecsficu2009} and loss measurements based on coplanar waveguide resonators~\cite{Pappas2011}. In addition, comparing microwave loss in plasma-grown aluminum oxide with loss in films fabricated using other methods, \emph{i.e.} low pressure oxidation~\cite{Martinis2005, Kim2008, Lupaifmmodemboxcselsecsficu2009} (the standard method for fabrication of Josephson junctions) and deposition of aluminum in a reactive oxygen atmosphere~\cite{Pappas2011}, is relevant for understanding this material. We note that an aluminum oxide growth method similar to the method reported here was demonstrated in ref.~\onlinecite{Heij_2001_PhDthesis}. However, the temperature and power dependence of microwave loss in plasma grown oxide has not been previously characterized. The small thickness and the robustness of the aluminum oxide layers obtained in our work are relevant for microwave circuits with compact capacitors, especially in applications where relatively large loss can be tolerated, such as parametric amplifiers~\cite{Hatridge2011}.

\begin{figure}
  \includegraphics[width=8.5cm]{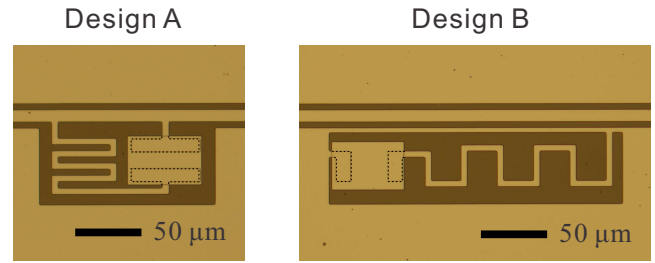}
  \caption{Optical microscope images of overlap-capacitor resonators with overcoupled~(Design A) and undercoupled~(Design B) designs. The dashed lines indicate the contour of the capacitor bottom plates.}\label{fig:resonator}
\end{figure}

We study dielectric loss using superconducting LC resonators formed of a meander inductor and an overlap capacitor containing aluminum oxide as a dielectric (see Fig.~\ref{fig:resonator}). The resonators are fabricated on oxidized high-resistivity silicon substrates. The fabrication process has three steps: (1) electron-beam deposition of an 100 nm aluminum layer through a mask defined by optical lithography, followed by lift-off; (2) plasma oxidation, to form an aluminum oxide layer on top of the aluminum film obtained in step 1; (3) electron-beam deposition of an 125 nm aluminum layer through an electron-beam defined mask, followed by lift-off. The plasma oxidation is performed at 200~$^\circ$C at an oxygen pressure of 0.2~mbar for 5 minutes. With these oxidation parameters, we find a characteristic capacitance of 16~fF/$\mu$m$^2$ at room  temperature. Electron-beam deposition and plasma oxidation are performed using a Plassys MEB 550S system.

We fabricated two nominally identical chips. Each chip contains a few lumped resonators, with different resonance frequencies, coupled to a common coplanar waveguide transmission line. We used two designs, A and B (see Fig.~\ref{fig:resonator}), which are overcoupled and undercoupled respectively. In addition, each chip contains one resonator based on an interdigital capacitor, which is used to determine an upper bound on other sources of loss. The parameters of the resonators are shown in Table~\ref{tab:para}. Transmission measurements are performed in a dilution refrigerator, using the system described in Ref.~\onlinecite{Ong2012}. The input line has a total of 50~dB attenuation distributed at different temperature stages. The output line is equipped with two isolators and a high electron mobility transistor cryogenic amplifier with a noise temperature of 4~K. Measurements of the $S_{21}$ scattering parameter are performed using a vector network analyzer.

We use the method presented in Ref.~\onlinecite{Deng2013} to relate the the $S_{21}$ scattering parameter to the resonator quality factors. We have:
\be
S_{21}(\omega) = \frac{1+2i Q_i \frac{\omega - \omega_0}{\omega_0}}{1+\frac{Q_i}{Q_e}+i\frac{Q_i}{Q_\alpha}+2i Q_i \frac{\omega-\omega_0}{\omega_0}}, \label{eq:S21}
\ee
where $Q_i$ is the internal quality factor, $Q_e$ is the external quality factor, $\omega_0$ is the resonance frequency, and $Q_\alpha$ is an asymmetry parameter of the resonator. We fit Eq.~\eq{S21} to the $S_{21}$ data at each excitation voltage. In order to reliably determine the quality factor from the relatively noisy data at low excitation voltages, we first fit Eq.~\eq{S21} to the data at high excitation voltage to determine $Q_e$, $Q_\alpha$, and the transmission baseline for each resonator. These parameters are independent of the loss of the resonators. Then the low power $S_{21}$ data is fitted with only $Q_i$ and $\omega_0$ as free parameters. The quality factor $Q_i$ of overlap capacitor resonators is a measure of the aluminum oxide loss tangent. Indeed, the resonators with interdigital capacitors have quality factors in the $10^{5}$ range, which is two orders of magnitude higher than that of the overlap capacitor resonators. This shows that sources of loss other than the dielectric in the overlap capacitor are negligible. We determine the loss tangent as a function of the voltage across the capacitor. The capacitor voltage $V$ can be related to the forward propagating voltage $V_{\t{in}}^+$ at the input port using the formula
\be
V = V_{\t{in}}^+  \times \frac{\lambda}{1/Q_i+1/Q_e+i/Q_\alpha} \label{eq:V},
\ee
where $\lambda$ is a coupling factor. The coupling factor $\lambda \approx M\omega_0/Z_0$, where $M$ is the mutual inductance between the resonator and the transmission line. The mutual inductance $M$ is 100~pH and 18~pH in design A and B respectively, as determined using numerical simulations.

\begin{figure}
  \includegraphics[width=8.5cm]{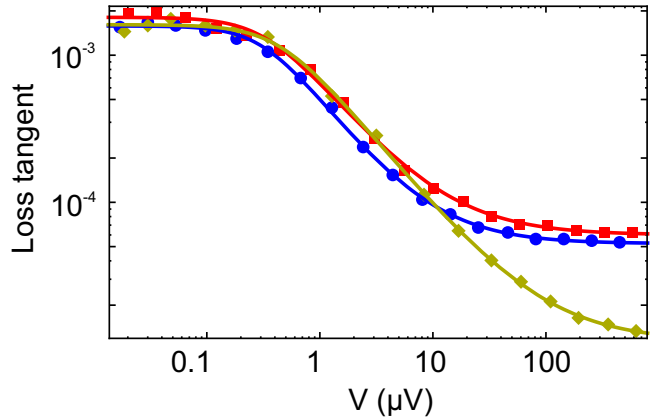}
  \caption{Loss tangent versus voltage on capacitor, for data of resonators on chip C1: LC1 at 4.9~GHz (blue dots), LC2 at 5.8~GHz (red squares), and LC3 at 6.2~GHz (yellow diamonds) and the fits with TLS model (Eq.~\eq{TLS}). The measurements are done at 54~mK.} \label{fig:power_dep}
\end{figure}

We first present the measurements of the dependence of dielectric loss on the voltage $V$. Fig.~\ref{fig:power_dep} shows the loss tangent versus the voltage on the overlap capacitors in different LC resonators at 54~mK. The loss saturates when the voltage on the capacitor corresponds to $<$0.1 photons stored in the resonator. The voltage dependence of the loss tangent is fitted to the two-level system (TLS) model\cite{Phillips1987, Martinis2005}:
\be
\frac{1}{Q_i} = \frac{1}{Q_{\t{offset}}}+\frac{\tan \delta}{\sqrt{1+(V/V_c)^{2}}},\label{eq:TLS}
\ee
where $\tan \delta$ is the loss tangent of the dielectric in the sub-single photon regime, $V_c$ is a critical voltage, and $Q_{\t{offset}}$ represents a voltage-independent loss likely due to quasiparticles excited by stray thermal radiation. Table~\ref{tab:para} is a summary of the measurements of different resonators on the two fabricated chips at 54~mK. The TLS loss model, given in Eq.~\eq{TLS}, fits all the data very well. We consistently obtain a loss tangent between $1.4\times 10^{-3}$ and $1.8\times 10^{-3}$ in the low voltage saturation regime.

\begin{table}[]
\begin{tabular}{p{0.8cm} p{1.5cm} p{1.0cm} p{1cm} p{1.2cm} p{1cm} p{1cm}}
  \hline \hline
  Chip & Resonator \& design & $f_0$ (GHz) & $Q_e$ & $\tan\delta$ $(\times10^{-3})$ & $Q_{\t{offset}}$ & $V_c (\mu V)$\\ \hline
  C1 & LC1, A & 4.91 & 336 & 1.54 & 19366 & 0.300\\
  C1 & LC2, A & 5.79 & 380 & 1.72 & 16245 & 0.371\\
  C1 & LC3, B & 6.16 & 3886 & 1.57 & 79194 & 0.534\\
  C2 & LC1, A & 5.03 & 310 & 1.75 & 38462 & 0.256\\
  C2 & LC2, A & 5.75 & 369 & 1.52 & 35397 & 0.424\\
  C2 & LC3, B & 6.08 & 3939 & 1.40 & 16805 & 0.399\\
  \hline \hline
\end{tabular}
\caption{\label{tab:para}Summary of the measurement results of resonators on two different chips at 54~mK.}
\end{table}

\begin{figure}
  \includegraphics[width=8.5cm]{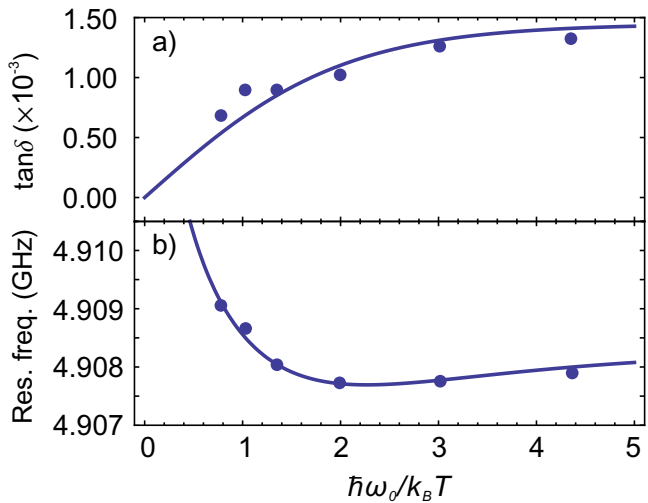}
  \caption{Temperature dependent data of the sub-single photon dielectric loss (a) and the frequency shift (b) of the resonator C1-LC1. The resonance frequency in (b) is obtained at $V \approx 0.03\, \mu $V. The solid lines are the fits to the TLS model.}\label{fig:temp_dep}
\end{figure}

We next discuss the temperature dependence of the dielectric loss. The voltage-dependent dielectric loss of resonator LC1 on C1 is measured at different temperatures. The data at each temperature is fitted with Eq.~\eq{TLS}, which yields the low-power loss tangent $\tan \delta$. The TLS model predicts the following depenence of the loss tangent on temperature $T$:
\be
\tan \delta = \tan \delta_0 \times \tanh \left( \frac{\hbar\omega_0}{2k_B T} \right) \label{eq:TLS_T},
\ee
with $\tan \delta_0$ the intrinsic TLS loss tangent. In Fig.~\ref{fig:temp_dep}(a), we plot $\tan \delta$ versus temperature, and also a fit of the model in Eq.~\eq{TLS_T}. The fit yields $\tan \delta_0 = 1.5\times 10^{-3}$. In Fig.~\ref{fig:temp_dep}(b), we show the resonance frequency at low power versus temperature. We fit this data with a prediction based on TLS theory\cite{Phillips1987}:
\be
f_0(T) = f_{0,i}\left(1+p \left(\t{Re}\Psi \left(\frac{\hbar\omega_0}{2\pi i k_B T}+\frac{1}{2}\right)-\ln \frac{\hbar\omega_0}{k_B T}\right)\right),
\ee
with $f_{0,i}$ and $p$ as fit parameters. TLS theory gives $\tan\delta_0 = \pi p$, which provides an alternative to determining $\tan \delta_0$; we find a value $\tan \delta_0= 2\times 10^{-3}$ from the fit, in good agreement with the determination based on the variation of low-power loss with temperature.

In the following we discuss the chemical composition of the aluminum oxide films. The films are characterized using X-ray photoelectron spectroscopy (XPS). The plasma oxide has an oxygen-to-aluminum atomic ratio of $1.69$. The binding energy (BE) of oxygen $1s$ electron is $531.3$~eV. The dependence of the XPS intensity signal on energy for the aluminum $2p$ electrons from the metal and the oxide layer are used to determine the oxide thickness, based on Strohmeier Equation~\cite{Strohmeier1990}. We determine a thickness of $5.0$~nm. For comparison, we measure the O-Al atomic ratio for aluminum oxide films grown by atomic layer deposition (ALD) and exposure to air. We find the ratios $1.52$ and $2.08$ respectively. The BEs of the oxygen $1s$ electron of the ALD oxide and the native oxide are $531.3$~eV and $532.3$~eV respectively. With the $5$~nm thickness of the plasma oxide and the 16~fF/$\mu$m$^2$ capacitance density obtained from room-temperature measurements of these capacitors, we deduce that the relative permittivity $\epsilon_r = 9.0$. In spite of the small thickness and the high capacitance density, tested plasma oxide capacitors with capacitance ranging from $1.5$~pF to $10$~pF have a breakdown voltage greater than $10$~V at room temperature. The process yield is very high: the specific capacitance of more than 40 capacitors, with areas between 100 and 600~$\mu$m$^2$, fabricated in different runs, had a spread of less than 1\%. The 6 resonators measured at low temperatures have resonance frequencies in good agreement with the results of microwave simulations assuming the capacitance as determined at room temperature using a low-frequency measurement.

In conclusion, we fabricated aluminum oxide layers using plasma oxidation. This is a high yield process, producing a robust and thin oxide layer. We performed a detailed characterization of the temperature dependence of microwave loss. The results are in good agreement with the tunneling two-level system model. The values of the loss tangent are comparable with values obtained for other types of amorphous aluminum oxide~\cite{Martinis2005, Pappas2011}. Our results are relevant for the field of superconducting quantum devices, where aluminum oxide is the tunnel barrier of Josephson junctions used for quantum bits. Plasma oxidation is suitable for compact superconducting resonators where loss is not critical, in applications such as parameteric amplifiers.

We are grateful to Patrick Smutek for help and advice concerning the plasma oxidation process. We also thank Liyan Zhao for advice on XPS measurements, Mustafa Bal for the help on sample fabrications, and Jean-Luc Orgiazzi and Florian Ong for the help with the low-temperature experiments.
This work was supported by NSERC, CFI, Industry Canada, Ministry of Research and Innovation and CMC Microsystems. C.D. was supported by an Ontario Graduate Scholarship. A.L. was supported by a Sloan Fellowship and an Early Researcher Award.

\end{document}